\documentclass{article}
\usepackage{spconf,amsmath,amssymb,graphicx}
\usepackage{float, boldline, multirow}

\usepackage{ragged2e}

\AtBeginDocument{
    \setlength{\abovedisplayskip}{3.25pt}
    \setlength{\belowdisplayskip}{3.25pt}
    \setlength{\abovedisplayshortskip}{0pt}
    \setlength{\belowdisplayshortskip}{0pt}
}

\title{Classification-Oriented Semantic Wireless Communications} 
\name{Emrecan Kutay and Aylin Yener}
\address{Dept. of Electrical and Computer Engineering, The Ohio State University}

\begin{document}
\maketitle

\begin{abstract} 
We propose semantic communication over wireless channels for various modalities, e.g., text and images, in a task-oriented communications setup where the task is classification. We present two approaches based on memory and learning. Both approaches rely on a pre-trained neural network to extract semantic information but differ in codebook construction. In the memory-based approach, we use semantic quantization and compression models, leveraging past source realizations as a codebook to eliminate the need for further training. In the learning-based approach, we use a semantic vector quantized autoencoder model that learns a codebook from scratch. Both are followed by a channel coder in order to reliably convey semantic information to the receiver (classifier) through the wireless medium. In addition to classification accuracy, we define system time efficiency as a new performance metric. Our results demonstrate that the proposed memory-based approach outperforms its learning-based counterpart with respect to system time efficiency while offering comparable accuracy to semantic agnostic conventional baselines. \end{abstract}


\vspace{-0.1in}
\section{INTRODUCTION}
\label{sec:intro}
The increasing prevalence of human-machine and machine-machine interactions has led to the emergence of resource-intensive applications, such as digital twins, autonomous vehicles, and intelligent environments \cite{autonomousvehicles,IoT_survey}.  
These applications typically have stringent latency requirements and need to send and receive vast amounts of data, which can overwhelm wireless links. Further, there can be extreme communication scenarios in next generation systems that aim to connect everything in the universe, where scarce resources must be very effectively managed.  Semantic communications \cite{guler2014semantic} is an emerging field that aims to reliably convey {\it meanings} of messages to the end-user, rather than reliably conveying the exact message sent \cite{Shannon}. Semantic communications in a sense extracts the useful information and as such can lead to resource savings, especially in the emerging paradigm of task-oriented communications, i.e, when the receiver has a task to accomplish \cite{beyond_bits_overview}. 

Recent semantic communication models utilize neural networks (NN) to extract and convey the underlying meaning. For example, in \cite{zhijin_deepsc}, transformer NNs, along with dense layers, have been employed to obtain and transmit semantic information over noisy channels for text communication. For image transmission, reference \cite{gunduz_wireless_image} utilizes convolutional layers to directly map images to channel symbols, subsequently reconstructing them at the receiver. Motivated by the inherent end-to-end differentiability, existing deep learning (DL)-based models, not limited to the references mentioned, follow an end-to-end training paradigm. However, the associated cost of this training is substantial in terms of time, energy, and computations \cite{semantic_com_challenges}. Even when assuming a manageable training cost, such an approach can limit generalizability, as it customizes models for specific contexts, inherently linked to the characteristics of the task and training dataset. In real-life use cases with highly dynamic or extreme environments, we may not be able to afford such customization and may have to think of alternatives \cite{imperfect_opt}. References \cite{zhijin_deepsc} and \cite{channel_transfer} address this issue partially utilizing a transfer learning approach, but do not eliminate the training overhead completely.

In machine learning (ML) community, utilization of pre-trained neural networks in inverse problems has shown great potential \cite{ilo}. Reference \cite{compressed_sensing_generative_models} focused on the sparse recovery problem of images and demonstrated that the use of a pre-trained generator network can outperform the reconstruction performance of the traditional Lasso algorithm, without requiring an additional training phase on the NN. Reference \cite{model_based_dl} illustrates that, instead of using a pure DL structure, combining it with conventional blocks can establish a more interpretable setup.  This approach addresses a characteristic challenge of DL models, which are often treated as black boxes. The same challenge is also present in recent semantic communication models, due to their reliance on DL, where researchers are seeking interpretability \cite{semantic_encoder}.

In this paper, inspired by these recent advances in the ML community, we utilize pre-trained models for semantic communications over wireless channels for text and image when we have a task-oriented scenario with the task at hand at the receiver is classification. Specifically, building on our recent work for semantic text compression \cite{emrecan_icc_23}, we utilize two approaches based on memory and learning for the semantic communication of sources, applicable for text and image modalities. Our primary goal is to convey the underlying meaning rather than the exact representation for message classification at the destination. While classification accuracy remains a metric we evaluate, we also define a new performance metric that measures the amount of tasks that can be classified in a given time budget. Termed {\it system time efficiency}, this metric highlights the savings that could be obtained by removing the training overhead. Our results demonstrate that memory-based models can deduce the semantic structure of a source without imposing any assumptions and without necessitating a training phase. Hence, they achieve higher system time efficiency and greater accuracy, particularly in scenarios with limited training data, which poses a significant challenge in DL models. Memory-based models provide a more explainable structure compared to the learning-based model. 

\vspace{-0.085in}
\section{SYSTEM MODEL}
\label{sec:system_model}
\vspace{-0.085in}
We consider task-oriented semantic communications, where transmitted image and text messages are classified at the receiver’s end. That is a classifier is only available at the receiver and not at the transmitter, necessitating source transmissions. Such a scenario, could be in an extreme environment, e.g., a disaster rescue scenario where conveying of certain information from a wireless device/sensor can help the receiver determine, location or the state of the survivors. We consider additive white Gaussian noise (AWGN) and Rayleigh fading channels. 
The received signal $ \mathbf{Y} \in \mathbb{C}^B$ is 
\begin{equation}
    \mathbf{Y} = h\mathbf{X} + \mathbf{Z} 
\end{equation}
\noindent
where $\mathbf{X} \in \mathbb{C}^B$ is the channel input, $\mathbf{Z}$ is Gaussian noise, $\mathbb{C}\mathcal{N}(0, \sigma_Z^2\mathbf{I}_{B x B})$, and $h$ represents the channel gain. We set $h=1$ for AWGN channels, whereas  $h \sim \mathbb{C}\mathcal{N}(0, \sigma_h^2)$ for Rayleigh fading channels, where we set $\sigma_h^2=1$. We assume a unit power constraint on the channel input. These parameter settings result in a signal-to-noise ratio (SNR) of $-10\log_{10}({\sigma_Z^2})$ dB for both channel models. To emulate different channel SNR values, we adjust the noise variance, $\sigma_Z^2$. For Rayleigh fading, we employ transmit power control (channel inversion). For our simulations, we assume perfect channel state information.

\vspace{-0.085in}
\subsection{Classification Block}
\label{sec:classification_block}
\vspace{-0.08in}
This block performs classification at the receiver by reconstructed semantic feature vectors (embeddings), denoted as $\hat{Q} \in \mathbb{R}^p$. To achieve this, we have designed a shallow NN structure with three layers that consist of the number of hidden units: 128, 32, and $n_{\text{class}}$, respectively. Here, $n_{\text{class}}$ denotes the number of classes in the dataset. We keep the structure simple as this block is not the primary focus of this paper. The receiver dataset, including training embeddings $Q_{\text{train}}$ with associated labels $y_{\text{train}}$, is employed for training. 

\vspace{-0.08in}
\section{PROPOSED MODELS}
\label{sec:proposed_approaches}
\vspace{-0.08in}
In both our approaches, we utilize sentence-bidirectional encoder representations from transformers (SBERT) with the OpenAI CLIP NN to extract semantic information from both image and text sources \cite{clip_embed}. SBERT provides semantic information in the form of embeddings, denoted as $Q \in \mathbb{R}^p$, based on the principle that semantically similar messages are closely mapped, while those with dissimilar meanings are represented as distant real vectors. This is achieved through fine-tuning pre-trained transformers via classification, regression, and triplet objective functions. This process enables these transformers to function as semantic extraction modules without requiring a further training phase \cite{sentence-bert}.

We employ the same dataset $Q_{pre}^N$ for all models, storing embeddings of previous messages in an \(N \times p\) matrix, where \(N\), and \(p\) denote the number and dimension of embeddings. This constraint, influenced by \(N\), governs the amount of available data. Building upon our prior research on memory-based semantic text compression \cite{emrecan_icc_23}, we extend our study to encompass multiple modalities and investigate the most efficient utilization of the available data, $Q_{pre}^N$, by comparing it with a learning-based approach over wireless channels.
 
\vspace{-0.05in}
\subsection{Semantic Quantization}
\label{sec:semantic_quantization}
\begin{figure}
\centering
\includegraphics[width=2.625in]{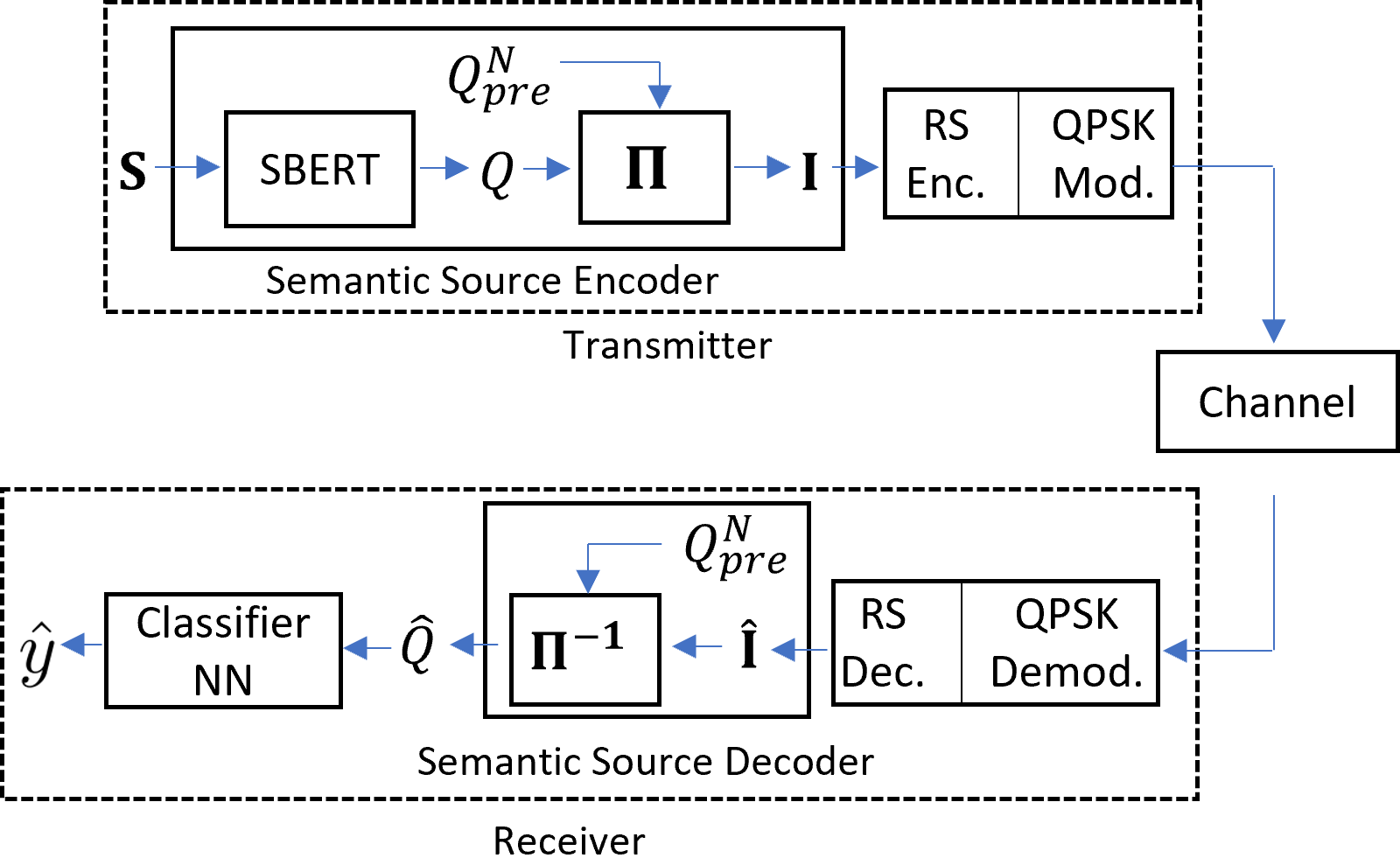}
 \vspace{-0.12in}
\caption{\small The model of memory-based semantic quantization.}
\label{fig:semantic_quantization_model} 
 \vspace{-0.1in}
\end{figure}

\vspace{-0.05in}
Source semantic structure can be discerned through prior realizations \cite{emrecan_icc_23}. In semantic quantization, we utilize $Q_{pre}^N$ as a codebook directly. As communication occurs over these codewords only (where source embeddings are quantized to one of the codewords), semantic distortion is introduced. We denote this semantic distortion by $\delta(.,.)$.
When we assign an embedding $\Tilde{Q}$ from $Q_{pre}^N$ in place of actual $Q$ for the message \textbf{s}, we have
\begin{equation}
    \label{semantic_distortion}
    \delta(Q, \Tilde{Q}) = \rVert Q - \Tilde{Q} \lVert_2
\end{equation} 
We obtain the codebook index by semantic index assignment \begin{math} \mathbf{\Pi}_{Q_{pre}^N}( . )\end{math} as follows.
\begin{equation}
    \label{index_assignment}
    \mathbf{\Pi}_{Q_{pre}^N}( Q \mathbf{)} = \underset{i' s.t. 0\le i' < N}{\mathrm{argmin}} \delta\Bigl(Q, Q_{pre}^N(i')\Bigl)
\end{equation}
Here, $Q_{pre}^N(i')$ denotes the $i'^{th}$ embedding. In fig. \ref{fig:semantic_quantization_model}, we select $\tilde{Q}$ using (\ref{index_assignment}), convert its index \textbf{I} to binary, and perform Reed-Solomon (RS) coding (rate $25/31$) with QPSK modulation. In this setup, accessing message contents corresponding to the embeddings in $Q_{pre}^N$ establishes an interpretable system. This facilitates examining the human-understandable content of assigned codewords in the form of text or image. 

\vspace{-0.05in}

\vspace{-0.05in}
\subsection{Semantic Compression}
\label{sec:semantic_compression}
\vspace{-0.05in}
\begin{figure}
\centering
\includegraphics[width=2.625in]{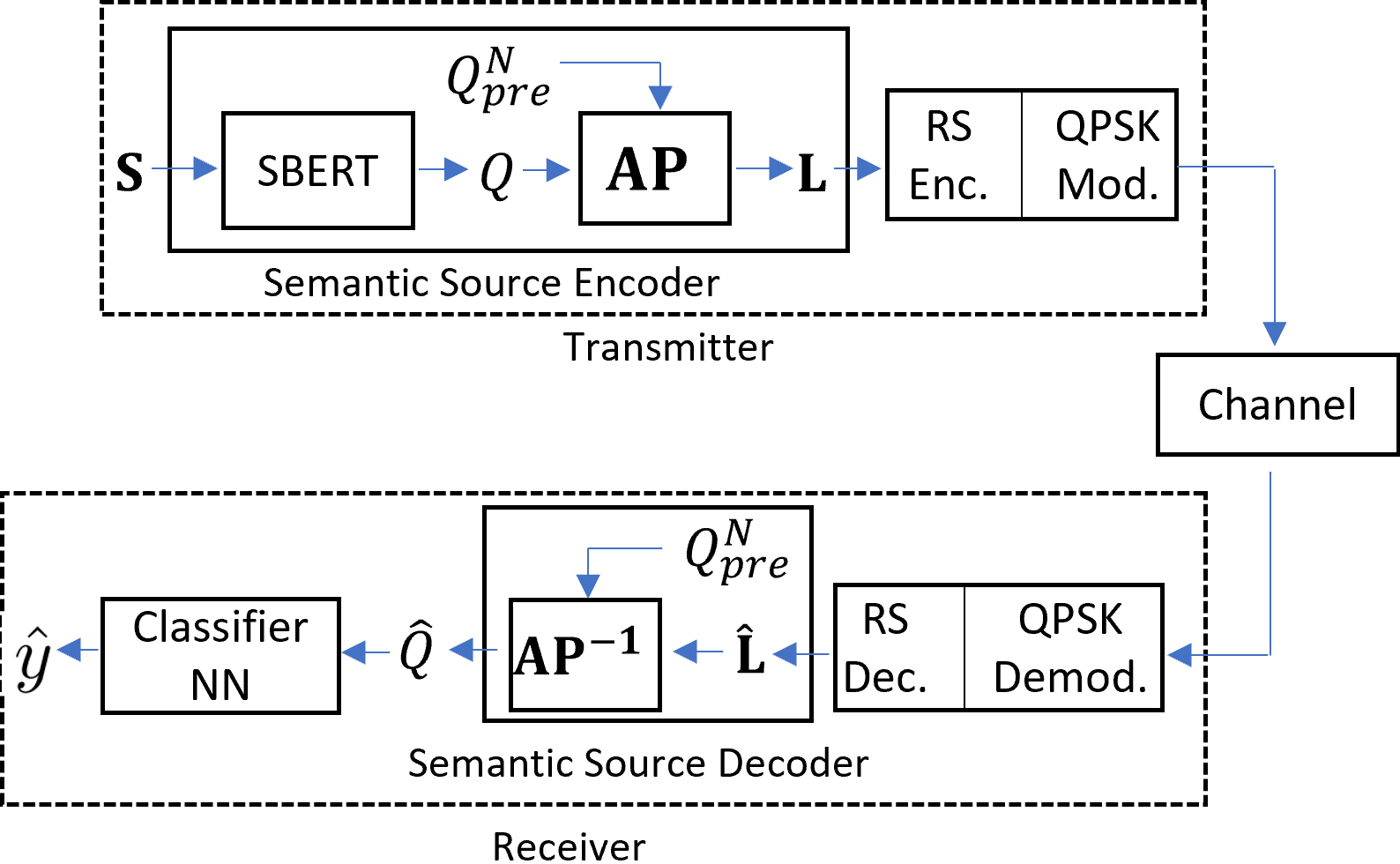}
 \vspace{-0.12in}
\caption{\small The model of memory-based semantic compression.}
\label{fig:semantic_compression_model} 
 \vspace{-0.1in}
\end{figure}

We present the semantic compression model building on semantic quantization. Although larger codebooks (\(N\)) improve (semantic) quantization resolution, they can lead to high storage and communication costs, necessitating more resources. To address this problem, we cluster embeddings in $Q_{pre}^N$ and using cluster centroids as codewords instead of individual elements in $Q_{pre}^N$ \cite{emrecan_icc_23}. We utilize the affinity propagation (AP) clustering algorithm as was done in \cite{emrecan_icc_23}, known for its good performance in minimizing Euclidean distances within clusters, aligning with equation \ref{semantic_distortion} \cite{affinity_prop}. Notably, this algorithm determines the number of clusters through its message-passing algorithm, rather than imposing fixed number of clusters. Since centroids are chosen from input data points by AP, the explainability of semantic quantization is retained. As depicted in fig. \ref{fig:semantic_compression_model}, after obtaining the corresponding cluster label \textbf{L}, we employ the same digital communication scheme with the same parameters as in section \ref{sec:semantic_quantization}.

\vspace{-0.05in}
\subsection{Semantic Vector Quantized Autoencoder}
\label{sec:vector_quantized_autoencoder}
\begin{figure}
\centering
\includegraphics[width=2.625in]{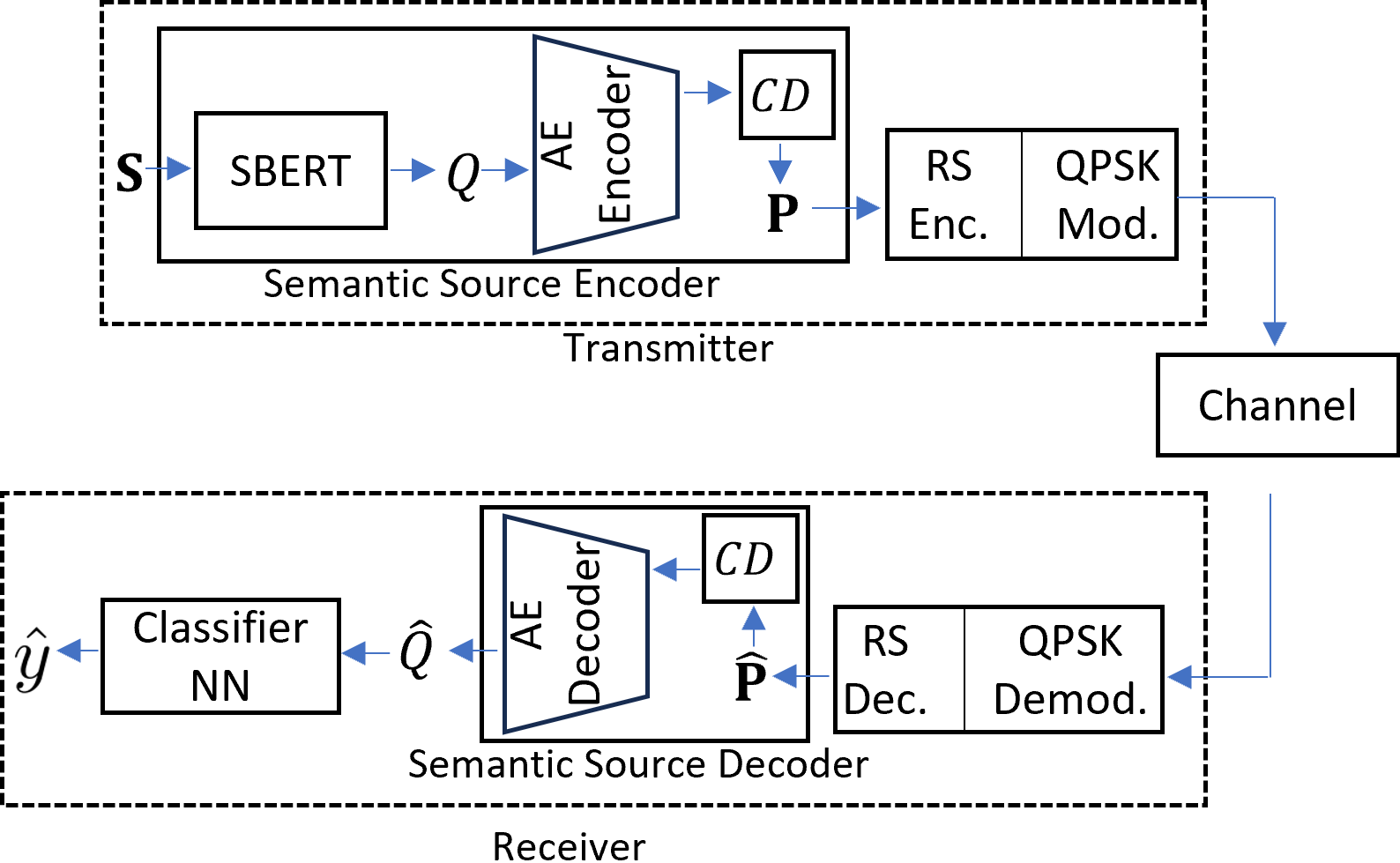}
 \vspace{-0.12in}
\caption{\small  The model of learning-based semantic VQ AE.}
\label{fig:semantic_vq_ae} 
 \vspace{-0.1in}
\end{figure}

\vspace{-0.05in}
We introduce a semantic vector quantized autoencoder model (VQ AE), inspired by the concept of vector quantized variational autoencoders \cite{vq_vae} and their recent applications in semantic image communication \cite{robust_vq_vae}. In contrast to the memory-based models, we utilize $Q_{pre}^N$ to train an autoencoder and a codebook, denoted as $CD$, providing discrete representations for given embeddings. This is achieved through vector quantization in the latent space of the encoder. We assign an element from the learned $CD$, minimizing $L2$ norm with the latent signal, in accordance with equation \ref{semantic_distortion}. Subsequently, the resulting $CD$ index \textbf{P} is converted to binary format and transmitted with the same digital communication setup as memory-based models, depicted in fig. \ref{fig:semantic_vq_ae}. We note that this model inherits data-driven characteristics observed in existing semantic communication models.

In designing the autoencoder, we propose an $\alpha$-scaled NN comprising fully connected (FC) layers to preserve information within the obtained embeddings while projecting them into a lower-dimensional latent space for quantization. For a given embedding dimension $p$, we perform mapping to an $\alpha^j$-dimensional space in the first FC layer of the encoder, where $j$ is the largest number such that $\alpha^j < p$. Subsequently, we iteratively scale the dimension down by ${\alpha}^{-1}$ in each layer until $\alpha^{\Tilde{j}}$, where $\Tilde{j}$ is the smallest number such that $K < \alpha^{\Tilde{j}}$. In the last FC layer, we directly map to $\mathbb{R}^K$, where $K$ denotes the specified latent space dimension. On the decoder side, the layers in the encoder are transposed to reconstruct the original embeddings given the codebook elements. For training, we follow the training algorithm of VQ VAE by setting the objective function as follows \cite{vq_vae}.
\begin{equation}
\label{vq_vae_objective}
    L = ||Q -\hat{Q}||_2^2 + ||sg[z(Q)] - e||_2^2 + \beta ||z(Q) - sg[e]||_2^2
\end{equation}
\noindent
Here, $z(Q)$ denotes the encoder output, $e$ represents the codebook element, $\beta$ is the commitment parameter, and $sg[.]$ is the stop gradient operation, ensuring that its elements are not updated by having zero partial derivatives. 

\vspace{-0.05in}
\subsection{Performance Metric}
\label{sec:performance_metric}
\vspace{-0.05in}
We propose  new metric  we term {\it system time efficiency}, denoted as $\eta_T(\Tilde{T})$, to evaluate the number of correct task executions (classifications) that proposed models can achieve within a given time budget $\Tilde{T}$. For a model that achieves $U$ correct classifications per second during the communication phase and requires $T_{train}$ seconds for the training phase, $\eta_T(\Tilde{T})$ is obtained as follows. 
\begin{equation}
    \label{eq:time_eff_metric}
    \eta_T (\Tilde{T}) = (\Tilde{T}-T_{train}) * U
\end{equation}

\vspace{-0.05in}
\section{RESULTS}
\label{sec:results}
\vspace{-0.05in}
We have conducted experiments using two datasets: i) AG's News, encompassing articles from four topics \cite{text_classification_dataset_lecun}; ii) STL 10, consisting of $96\times96$ RGB images belonging to ten object classes \cite{stl_10}. For the AG's News, we have allocated 10000 samples from the training set for $Q_{pre}^N$ and 2000 samples from the test set for simulation, with the remaining training set used for the classifier NN. 
For STL 10, we have first merged the training and test sets, resulting in a set of 13000 samples. We then have allocated 1000 samples for $Q_{pre}^N$ and another 2000 samples for testing, while the remaining samples were used for the classifier NN.

We utilize negative Euclidean distance as the similarity metric for AP, in conjunction with equation \ref{semantic_distortion}. We have conducted a grid-search-based hyperparameter tuning for both the classifier and the learning-based semantic VQ AE, as outlined in table \ref{table:training_parameters}. For a fair comparison, we set the codebook size in the semantic VQ AE to the number of clusters generated by AP, as indicated in table \ref{results_table}. We set $\alpha$ to 4 and determine the optimal latent space dimension through grid search. We have observed optimal performance in 64 and 16-dimensional latent spaces for AG's News and STL 10, respectively. This difference arises from the fact that the STL 10 dataset contains only 1000 samples for training its autoencoder and codebook, while AG's News has 10000 samples, imposing a stringent constraint to simulate scarce data.

\renewcommand{\arraystretch}{1.15} 
\begin{table}
\centering
\caption{\small Training Parameters}
\begin{tabular}{|c|c|c|}
\hline
\textbf{Parameter} & \textbf{Class. Block} & \textbf{Sem. VQ AE}  \\ \hline
Batch Size  & \multicolumn{2}{c|}{128} \\ \hline
Num. Epoch & 15 & STL 10 - 50 \\ 
 &  & AG's News - 30 \\ \hline
LR Scheduler & \multicolumn{2}{c|}{Exponential LR} \\ \hline
$\gamma_{Scheduler}$ & 0.75 & 0.97  \\ \hline
Initial & 0.001 & STL 10 - 0.01 \\
LR& & AG's News - 0.005 \\ \hline
Optimizer & \multicolumn{2}{c|}{Adam ($\beta_{1,2} = 0.9, 0.999$)} \\ \hline
\end{tabular}
\label{table:training_parameters}
 \vspace{-0.075in}
\end{table}
\renewcommand{\arraystretch}{1} 

To demonstrate the resource savings, we compare also with semantic-agnostic baselines. For these we have utilized Huffman coding for text and JPEG2000 for image compression. We have implemented block coding for Huffman, considering a group of three characters as a symbol. For JPEG2000, we have employed the Python Pillow package and set the quality layers to 20. Since both algorithms provide exact reconstruction, we calculated accuracy through the exact message embeddings obtained from SBERT.

\renewcommand{\arraystretch}{1.1} 
 \begin{table}[t]
 \centering
 \caption{\small Source Compression and Accuracy Results}
 \label{results_table}
 \begin{tabular} {!{\vrule width2pt}c|c|c|c!{\vrule width2pt}}
 
  \hlineB{5}
 \textbf{Dataset} & \multicolumn{1}{c|}{\textbf{Model}}  & \multicolumn{1}{c|}{\textbf{Bits (\%)}} & \multicolumn{1}{c|}{\textbf{Acc. (\%)}} \\
  \hlineB{5}
 \parbox[t]{3mm}{\multirow{4}{*}{\rotatebox[origin=c]{90}{AG's}}}\parbox[t]{3mm}{\multirow{4}{*}{\rotatebox[origin=c]{90}{News}}} & Conv. (Huffman) & 1481777 & 92.00\\
  \cline{2-4}

 & Sem. Quan & 28000  & 89.15\\
 \cline{2-4}

 & Sem. Comp. 944* & 20000 & 86.30\\
 \cline{2-4}
 
  & Sem. VQ AE 944$^+$ & 20000 & 80.65\\
  \hlineB{5}

 \parbox[t]{3mm}{\multirow{4}{*}{\rotatebox[origin=c]{90}{STL}}}\parbox[t]{3mm}{\multirow{4}{*}{\rotatebox[origin=c]{90}{10}}} & Conv. (JPEG2000) & 22194760 & 98.55\\
  \cline{2-4}

 & Sem. Quan & 20000  & 97.45\\
 \cline{2-4}

 & Sem. Comp. 63* & 12000 & 97.10\\
 \cline{2-4}
 
  & Sem. VQ AE 63$^+$ & 12000 & 92.10\\
  \hlineB{5}
 
 \end{tabular}
 \vspace{-0.1in}
 \justify
 \small{* Number of clusters calculated by AP for the corresponding dataset.} \\
 \small{$^+$ Codebook size of semantic VQ AE for the corresponding dataset.}
 \vspace{-0.25in}
 \end{table}
\renewcommand{\arraystretch}{1} 

\begin{figure}[t]
\centering
\includegraphics[width=\columnwidth]{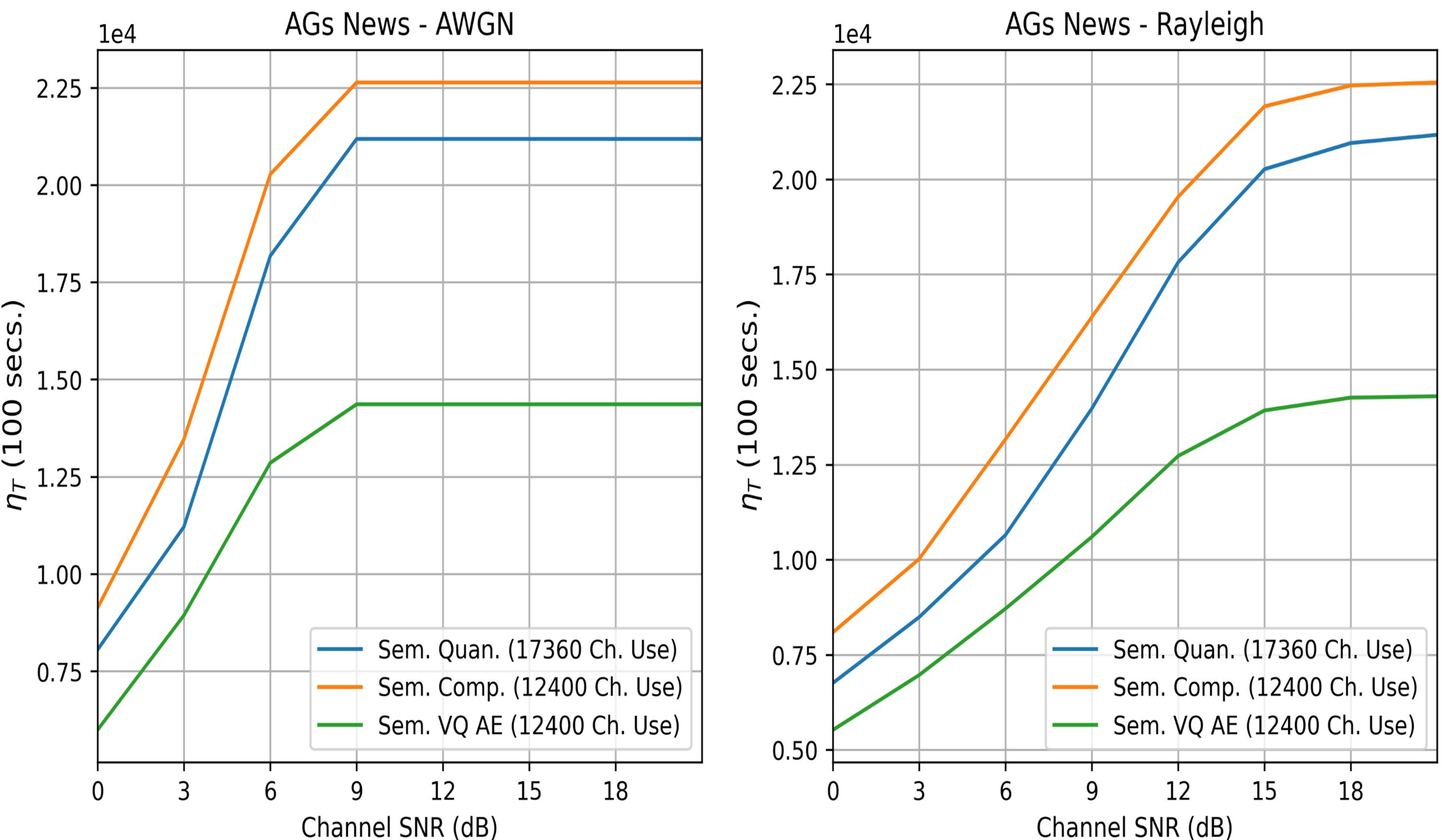}
\caption{\small System time efficiency results obtained for the AG's News dataset over AWGN and Rayleigh fading channels with a time budget of 100 seconds.}
\label{fig:ags_news_channel_results} 
 \vspace{-0.05in}
\end{figure}

\begin{figure}
\centering
\includegraphics[width=\columnwidth]{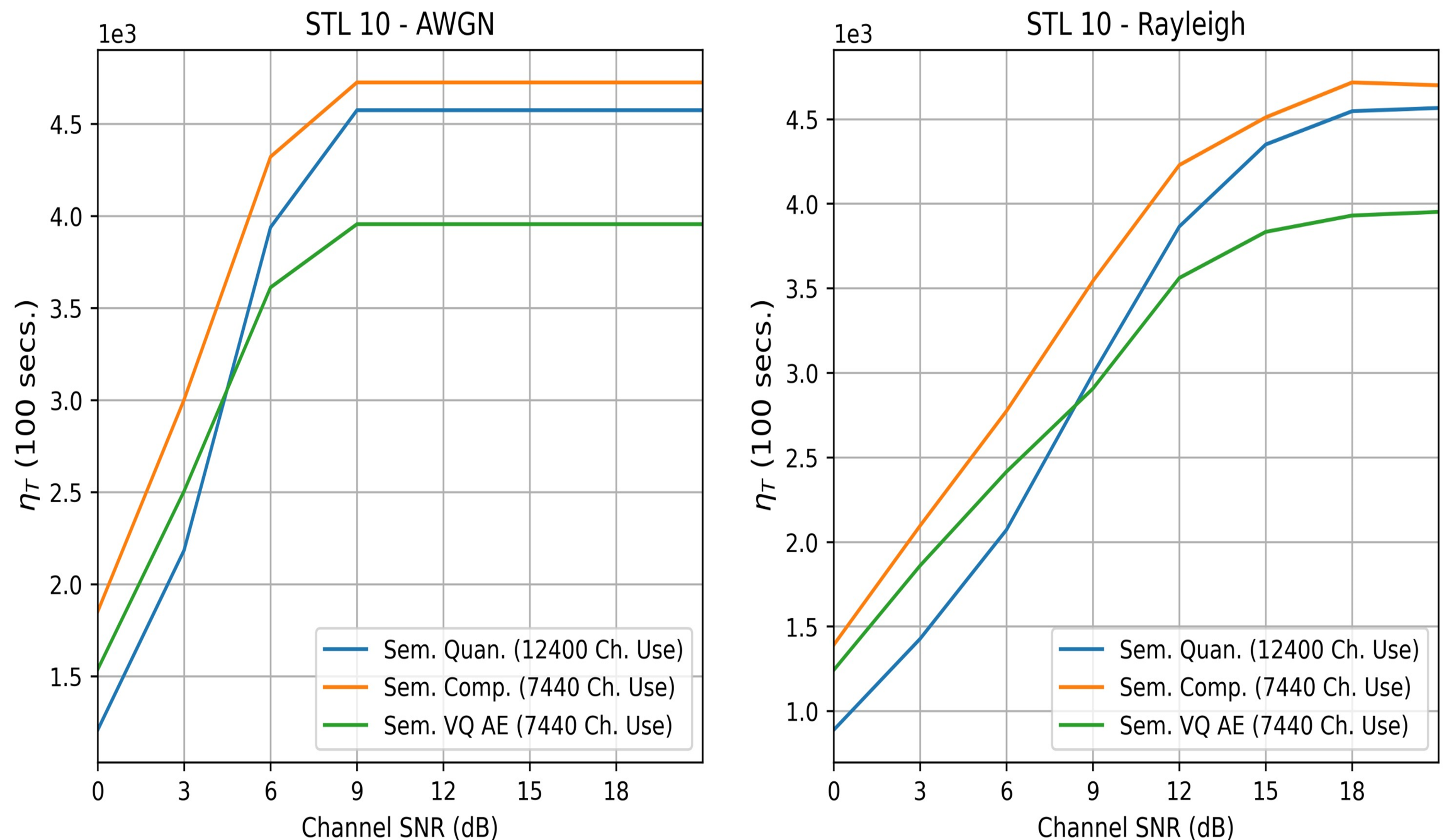}
\caption{\small System time efficiency results obtained for the STL 10 dataset over AWGN and Rayleigh fading channels with a time budget of 100 seconds.}
\label{fig:stl_10_channel_results} 
 \vspace{-0.15in}
\end{figure}

We start by assessing the efficacy of our semantic approaches in enhancing source compression efficiency. In table \ref{results_table}, it is evident that all semantic approaches demonstrate an order of magnitude improvement in both modalities. However, only semantic quantization and compression models maintain high levels of classification accuracy. This underscores the resilience of memory-based models in scenarios with less data, while learning-based semantic VQ AE model is notably affected. Expanding our evaluation to communication over wireless channels, memory-based models exhibit superior time efficiency compared to semantic VQ AE, emphasizing reduced energy and computation costs, as illustrated in figs. \ref{fig:ags_news_channel_results} and \ref{fig:stl_10_channel_results}. Elimination of training overhead has merit, particularly when evaluating performance on noisy channels within a limited time frame.

\vspace{-0.05in}
\section{CONCLUSION}
\label{sec:conclusion}
\vspace{-0.05in}

In this paper, we have semantic communication of text and images over wireless channels when the receiver has a classification task. We have utilized pre-trained models, introducing memory and learning-based approaches. Our results demonstrate that, especially in scenarios with limited data, direct utilization of available data as a codebook outperforms the recent semantic communication approach of training a DL model from scratch. Future work in this direction includes networked semantic communications in traditional as well as task-oriented scenarios.

\vfill\pagebreak

\bibliographystyle{IEEEbib}
\bibliography{refs}

\end{document}